# Reproducing a Security Risk Assessment Using Computer Aided Design

Avi Shaked[a]

[a]Department of Computer Science, University of Oxford,
Parks Road, Oxford,
OX1 3QD, UK
avi.shaked@cs.ox.ac.uk

## ABSTRACT

Security risk assessment is essential in establishing the trustworthiness and reliability of modern systems. While various security risk assessment approaches exist, prevalent applications are "pen and paper" implementations that – even if performed digitally using computers – remain prone to authoring mistakes and inconsistencies. Computer-aided design approaches can transform security risk assessments into more rigorous and sustainable efforts. This is of value to both industrial practitioners and researchers, who practice security risk assessments to reflect on systems' designs and to contribute to the discipline's state-of-the-art. In this article, we report the application of a model-based security design tool to reproduce a previously reported security assessment. The main contributions are: 1) an independent attempt to reproduce a refereed article describing a real security risk assessment of a system; 2) comparison of a new computer-aided application with a previous non-computer-aided application, based on a published, real-world case study; 3) a showcase for the potential advantages – for both practitioners and researchers – of using computer-aided design approaches to analyze reports and to assess systems.

## 1  Introduction

Reproducibility is the ability to reproduce results from available details (Rozier and Rozier, 2014). It is an important issue in research, where it supports rigorous validation of proposed methods and results as well as knowledge transfer and dissemination. Reproducibility is also of importance in industrial practices, whenever one wishes to apply engineering methods consistently and demonstrate high-maturity performance (Anda et al., 2009).

Security risk assessment is a process of identifying, analyzing, and evaluating potential threats to a system. Its primary objective is typically to assess the likelihood and impact of potential security incidents, supporting informed decision making about risk mitigation. Accordingly, security risk assessment incorporates aspects of threat modelling, design and risk management (Beling et al., 2019; De Rosa et al., 2022; Jbair et al., 2022). A recent systematic review on threat modelling identifies that it is *"a domain in which advancement is driven by both academia and industry"*, that it suffers from unsystematic applications, and that *"as an engineering discipline, is at a low level of maturity"* (Khalil et al., 2024).

Reproducing security risk assessments can contribute to improving the quality of the assessment reports as well as to validate the assessments' methods as engineering procedures, and, consequently, to advancing the engineering discipline. A recent systematic mapping study on security for systems of systems (SoS) reports that "*no single study fully described the SoS sufficiently to allow replication of the results obtained and support future research*"; and suggests that "*the design of next approaches and the validation of future work would benefit from security studies focusing on replicable processes rather than focusing on particular issues*" (Olivero et al., 2023). While there have been attempts to reproduce attacks (Lee et al., 2017), we are unaware of any scholarly efforts of reproducing a security risk assessment.

Here, we aim to reproduce a cybersecurity risk assessment for a specific SoS – an autonomous ship system – by Bolbot et al (Bolbot et al., 2020). We rely on the specific journal article – referred to as the original report – as the single source of information. We chose this specific risk assessment article for several main reasons: 1) the novel approach exercised in the original report is an end-to-end security risk assessment method, which relates to aspects of system design, threat modelling and risk management; 2) the exercised method – CYber-Risk Assessment for Marine Systems (CYRA-MS) – is highly representative of industrial security risk assessment efforts in which we participated, and therefore consider it an applied and practical method which can inform both researchers and practitioners about security risk assessment practices; 3) the original report is available under Common Creative license so that some of the original figures can be reproduced here for comparison and reflection throughout our reproduction.

We approach the reproduction of the original report using a computer-aided design methodology developed for supporting security engineering and assessment. The methodology is designed as a model-based approach and employs domain-specific knowledge, codified using a metamodel. Model-based and ontology-based approaches share metamodeling as a foundational concept (Wagner et al., 2023) and have been previously demonstrated for analyzing engineering artifacts and methods and improving reproducibility (Oliveira et al., 2023; Santana-Perez et al., 2017; Shaked and Reich, 2021; Zschaler and Polack, 2023).

Typically, model-based security assessment approaches lack support for risk management (Shaked, 2023). For example, several previous works use formal modelling – combining the informal Systems Theoretic Process Analysis (STPA) method and formal Event-B modelling – to elicit and analyze security requirements (Dghaym et al., 2021; Fathabadi et al., 2024), but without addressing risk management aspects. Furthermore, while novel, these formal modelling efforts offer specific approaches that do not necessarily allow for existing practices to be used and for previous reports – obtained using other practices – to be reproduced.

Here, we briefly discuss some approaches that address risk management and relate to model-based design. CORAS is an approach to risk analysis that claim to be model-driven (Lund et al., 2010). However, there is currently no tool support for exercising the approach end-to-end and for generating a complete information model. Additionally, CORAS does not inherently provide means to relate to the design of the system that is under assessment. Instead, an asset- and threat-focused perspective of the system is built using a threat diagram, and this does not communicate the composition of the system (Sahay et al., 2023). Jbair et al. offer an end-to-end threat modelling methodology, which addresses risk management aspects (Jbair et al., 2022). However, the supporting tool is not publicly available. The Reference Ontology for Security Engineering – ROSE – offers a formal ontology which integrates aspects of risk management and security engineering primarily for supporting risk treatment (Oliveira et al., 2022). ROSE does not include aspects of system composition, and some of the risk management aspects – such as representing risk tolerance – are not explicitly addressed. Also, a supporting tool – needed for instantiating the ROSE ontology and verifying designs according to the formal ontology's definitions – has yet to be developed. ThreMa is another formal ontology-based approach for threat modelling (De Rosa et al., 2022). While it allows depicting data flows and components of the analyzed system, it lacks support for understanding the hierarchical system composition and connectivity – which are captured using an informal diagram – and does not address risk management concepts. Yacraf (Yet Another Cybersecurity Risk Assessment Framework) is a metamodel-based cybersecurity risk assessment framework, which has yet to be implemented as a modelling tool (Ekstedt et al., 2023).

TRADES is a model-based security design and assessment methodology, supported by an open-source tool implementation (Shaked, 2023). It is designed as a meta-level methodology that could accommodate various security design and analysis methods, including existing practices. In this paper we use the TRADES modelling tool (henceforth "TRADES Tool") to reflect on the previously accomplished security assessment. The selection of a specific tool does not limit the generality of the findings, and we relate to this in the Discussion section.

Our reproducibility effort is aimed to highlight aspects of disciplined use of information models – particularly consistency and completeness – that can contribute to quality of engineering analysis and artifacts. Accordingly, this is not an end-to-end exercise nor a tutorial in cyber security risk assessment. We avoid specific discussion regarding cybersecurity aspects. This is done for three main reasons: 1) such discussions often require specific cybersecurity domain expertise – including cybersecurity for cyber-physical systems in the specific case – which is unreasonable to assume for the wider readership we wish to attract; 2) there is no consensus regarding pertinent knowledge, and, specifically, different contexts and applications of cybersecurity risk assessment may employ different definitions; 3) the cybersecurity threat landscape is dynamic, with attacker's capabilities gradually evolving. We encourage readers who are interested in the complementary aspects of cyber security assessment to learn the original report by Bolbot et al. We avoid repeating specific phases of the original assessment where cybersecurity knowledge is creatively exercised. We also avoid criticizing security-related assumptions or design choices. Still, we relate to the creative freedom and the rigor of the assessment whenever appropriate.

Instead, we highlight aspects of modelling and information structuring by attempting to reproduce specific information artifacts and their straightforward implications. By avoiding the application of domain-specific knowledge and focusing on broader information, methods and decision-making aspects, the findings are more likely to be transferable and adaptable across diverse engineering and managerial domains. In the recent systematic literature review of threat modelling, the authors identify that validation is hard to achieve due to the nature of the cybersecurity-related problem domain leading to inability to obtain pertinent data (Khalil et al., 2024). Our reproducibility effort can also be considered as a creative validation methodology to overcome this domain-inherent limitation.

For our reproducibility effort, we rely on the previously developed TRADES Tool ("TRADES Tool Repository," n.d.), without any extension; and seek to find out whether, how and to what extend the tool can accommodate a published, refereed cybersecurity risk assessment, which employs a novel assessment method. The developed analysis models are available online [A Github repository will be revealed after peer-review].

## 2   Security Risk Assessment Reproduction

This section provides the results of reproducing the original report (Bolbot et al., 2020). The application of the risk assessment method (CYRA-MS) to the specific system of interest (autonomous ship system) relies on a schematic diagram, depicting "*systems and equipment as well as their relevant interconnections and interactions.*" Figure 1 shows this schematic diagram.

Figure 2 shows a reproduction of the original report's schematic diagram in the form of a TRADES system design diagram. Apart from layout and elements resizing, the diagram in Figure 2 is fully a model-based representation, i.e., it is generated exclusively based on querying an underlying information model with respect to metamodel definitions (including coloring according to the link type legend). The only exception is the target arrowhead that is added to depict unidirectional physical interactions (TRADES links are currently all bi-directional by nature, with only *affect* relations – the TRADES equivalent of data flows – being depicted as unidirectional). To our knowledge, this is the first report to utilize the TRADES system design diagram and its underlying link type mechanism.

While reproducing the original schematic diagram, we note that the type of the link between *Ship System*s and *Other vessels* is unclear. Specifically, the notation used for expressing the link in the original report does not appear explicitly in the legend, opposed to the notation used for other links. It can be inferred that the type is a bi-directional form of "Physical interactions." However, the existence of the *Systems for communicating with other vessels* system block suggests that there is an undocumented communication with the other vessels by radio signals (i.e., electromagnetic waves). When contrasted with the use of the same notation between *Generators* and *Fuel system* and with the explicit mention of the *4G/5G* communication type (which is also by electromagnetic waves), we are unsure whether the original report considers radio signals as physical interaction.

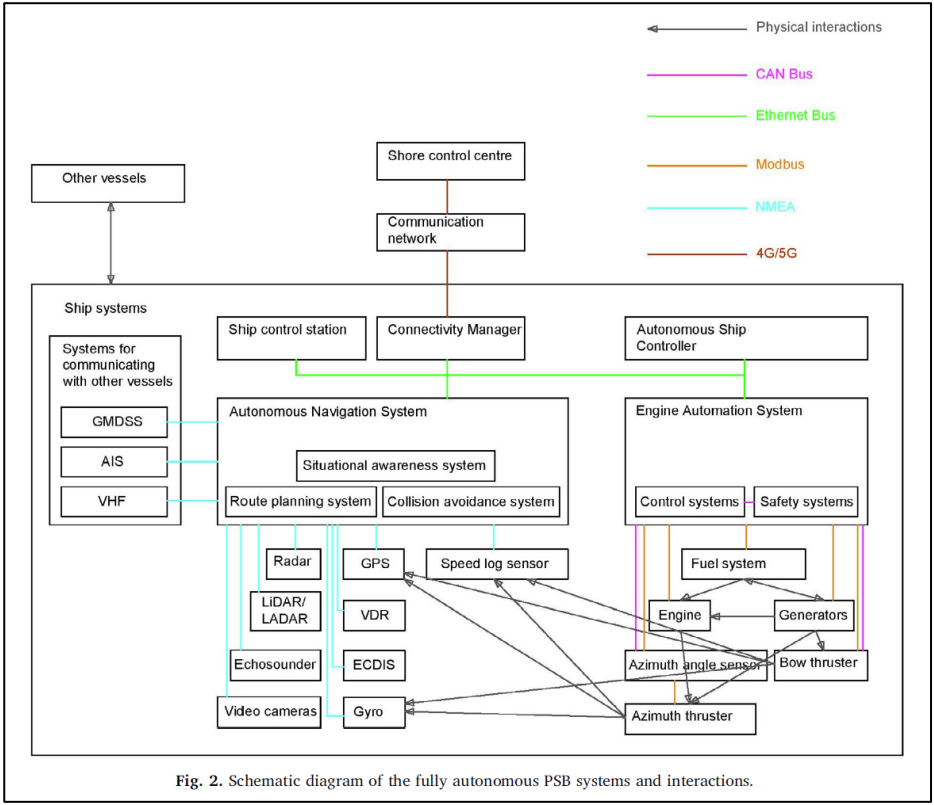

Fig. 2. Schematic diagram of the fully autonomous PSB systems and interactions.

**Figure 1: Schematic diagram as it appears in** (Bolbot et al., 2020) **(originally, Fig. 2)**

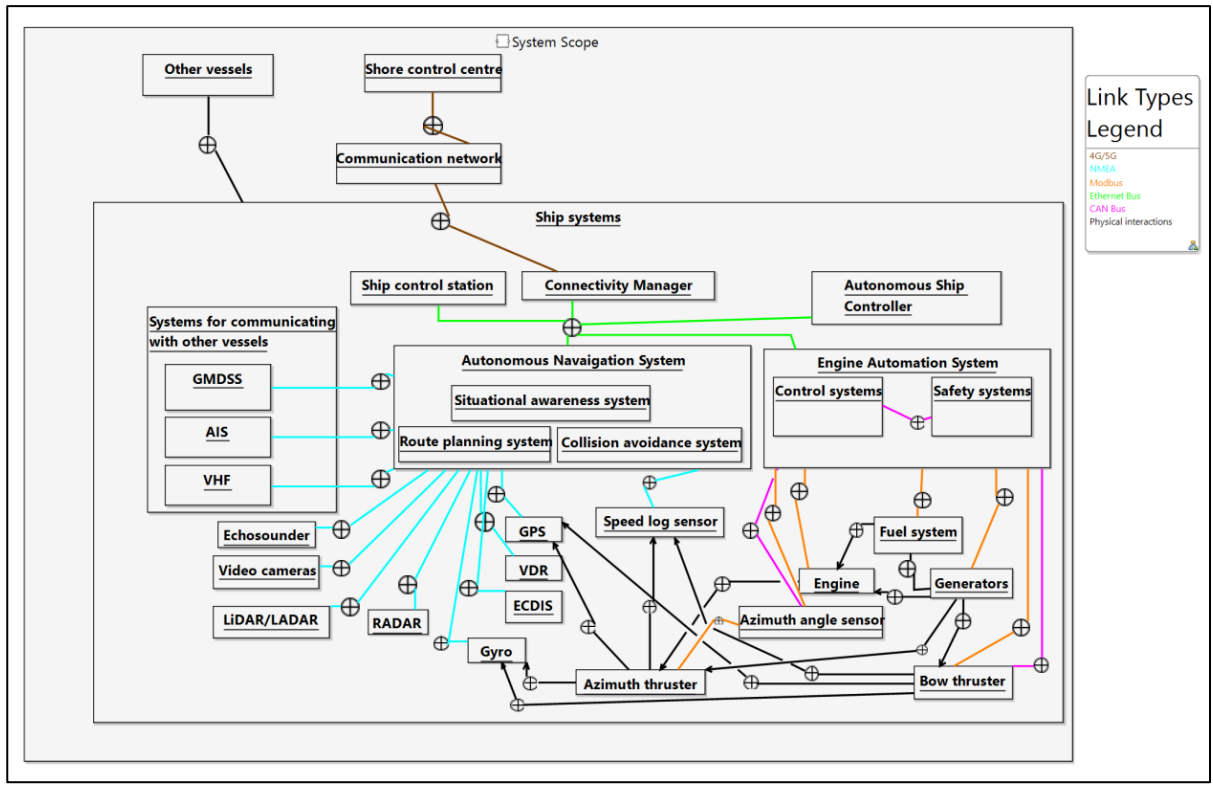

**Figure 2**: **Schematic diagram reproduced using TRADES**

Phase A of the assessment – *Preparation for analysis* – starts with step 1: *Systematic system analysis and review*. This step was reported to result in the artifacts shown in Figure 3 and Figure 4. The latter is only a representative example of such table-formed artifacts, but it is the only one revealed in the original report, somewhat limiting our reproduction.

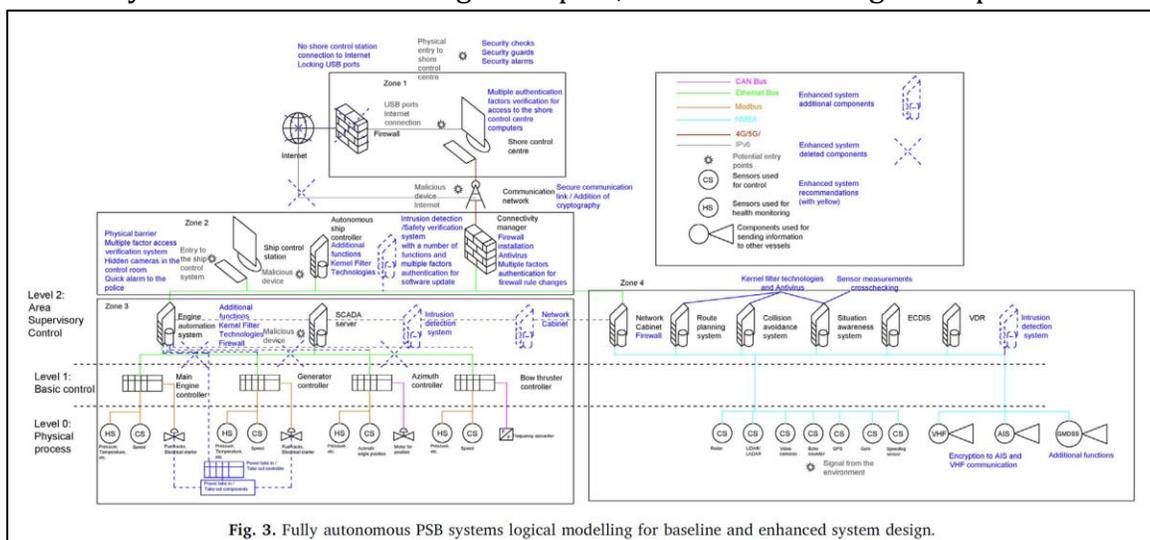

**Figure 3**: **System design as it appears in** (Bolbot et al., 2020) **(originally, Fig. 3)**

| Table 11 |||| 
| Investigated PSB selected components functionalities description. ||||
| Component | Functionalities | Data sent | Data received |
| --- | --- | --- | --- |
| Shore control centre | Monitoring of physical processes<br>Navigation control<br>Control over the ship in emergency/manoeuvring operating modes<br>Implementation of software updates | Control information for navigation<br>Selected route<br>Ship operating mode<br>Control status of equipment (on/off)<br>New software | Equipment health status<br>Equipment status (on/off, loads, position)<br>Images from cameras<br>Vessel position<br>VHF data<br>Traffic in the area<br>Radar, ECDIS information |

**Figure 4**: **Table analyzing functionality for a component, as it appears in** (Bolbot et al., 2020) **(originally, Table 11).**

By comparing the two original representations of the system design – Figure 1 and Figure 3 – we note the following discrepancies: 1) a single Ethernet bus (former) vs. two Ethernet buses (latter); 2) point to point NMEA communication links (former) vs. several NMEA buses (latter); no Internet connection (former) vs. newly disclosed Internet connection, typed IPv6 (latter).

We avoid reproducing the content of Figure 3, for three main reasons: 1) it is hard to determine which of the two original figures (Figure 1 & Figure 3) is a better representative of the real architecture; 2) whereas Figure 1 is explicitly mentioned as a basis for the assessment, reported to be integrated from multiple design artifact information sources, Figure 3 is a free-formed diagram, which involves a less disciplined and more complex notation and already involves some cybersecurity assessment elements (e.g., enhancement suggestions, as communicated by the legend). The nuanced details extend beyond the scope of this article, as we remind the reader that our focus is on the use of the information model and the consistent application of the reported method, and not on evaluating the exercised security domain knowledge; 3) the rest of the assessment can be communicated (to the reader) more easily using the initial representation (Figure 1). The only significant insight from Figure 3 that is employed by the original report's security assessment is the newly identified connection to the Internet. This is easily accommodated in a slight update to Figure 2. We, therefore, offer an updated TRADES representation of the architecture, in Figure 5, which communicates the existence of the *Internet* as a component within the scope of the risk assessment, as well as its IPv6-typed connection with the *Communication network* and the *Shore control centre* components.

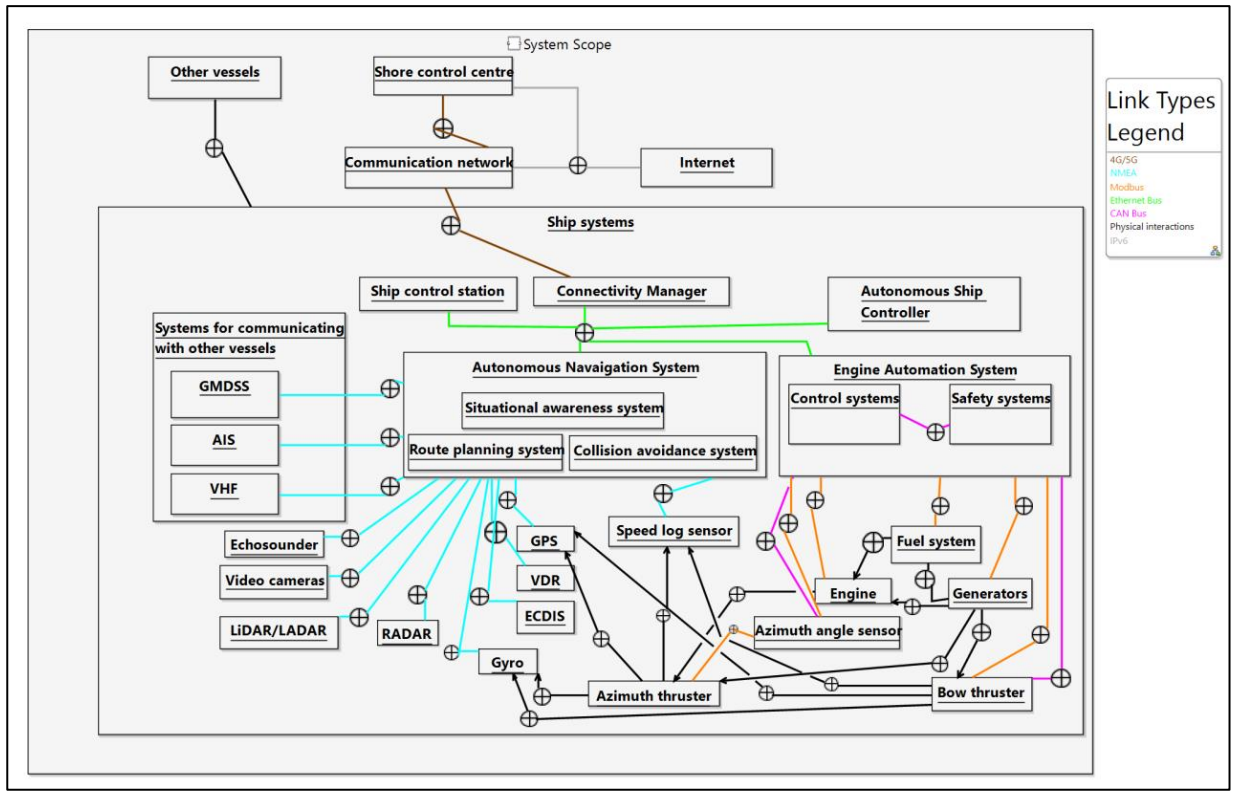

**Figure 5**: **Revised TRADES schematic diagram, including the Internet connection revealed in Figure 3.**

The additional resulting artifact available at this stage – Figure 4 – explicitly identifies sent and received data with respect to a specific component. However, when we try to model it using TRADES, a gap immediately appears: the other end of all the data flows is missing. Due to the specific scope of the case study in question, it is fairly intuitive to address this gap simply by stating that the missing end (source for received data and destination for sent data) is the *Ship systems*, i.e., the highest hierarchical level of organization of the systems within the autonomous ship platform. Figure 6 communicates this solving. The figure shows a TRADES design diagram generated from the information model, within the TRADES Tool modelling environment. Most of the lower-level elements of the *Ship systems* are hidden (using the diagram hide element tool), yet they remain part of the underlying information model. This diagrammatic representation allows specifying and viewing data flows (using the *Effect* tool from the *Palette* panel right of the diagram). In the figure, two specified data flows are shown using a thick, directed arrow between components (a standard tool-defined notation rooted in the TRADES methodology (Shaked, 2023)): one from *Shore control centre* to *Ship systems*, and another in the opposite direction. Each of these data flows convey specific data elements based on the original table artifact (Figure 4), with the entire list of unique data elements shown in the *Model Explorer* panel left of the diagram.

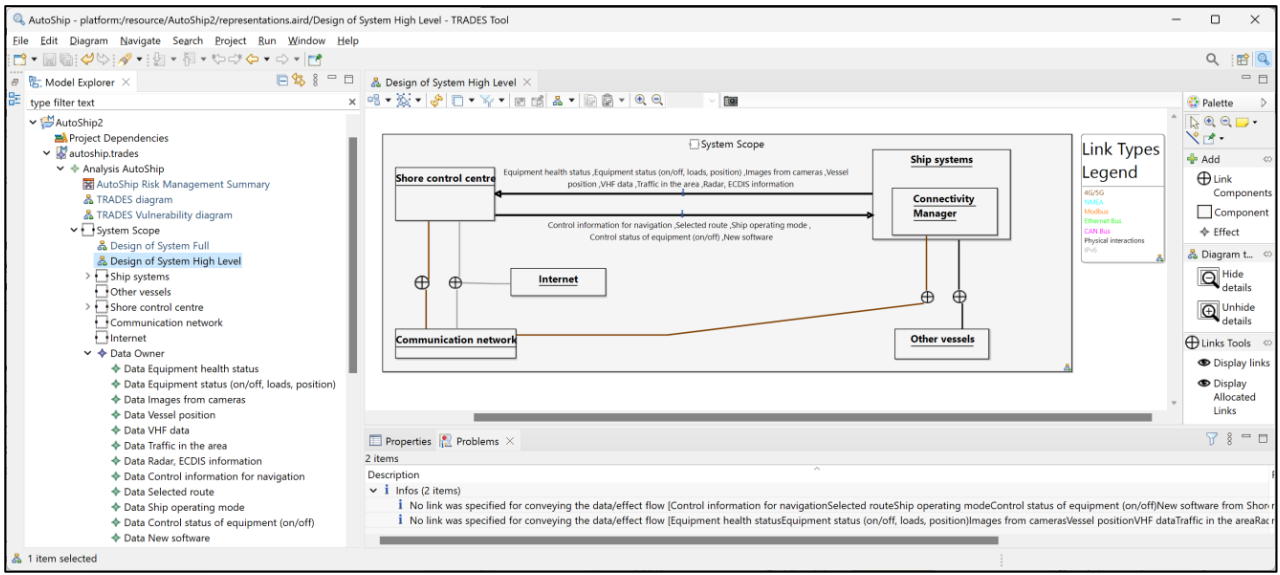

**Figure 6**: **Data flow depiction and analysis, within the TRADES Tool modelling environment.**

We wish to highlight that in a real risk assessment, more specific identities of the specific data flows' source/destination might be needed. The information about specific data flows' sources and destinations is not disclosed and cannot be reproduced here; yet, had it been provided we could have reproduced it by specifying the specific component as the source/destination for every data flow (in fact, TRADES also supports the creation of multi-hop data flows that relate to the same data entity).

At this stage, we can employ a TRADES diagram validation tool to see if our current codification of system design is complete. Specifically, the validation tool checks whether data flows are explicitly conveyed by existing links. Figure 6 also shows the results of applying this validation tool: the *Problems* panel, below the diagram, shows 2 *Info*-typed problem item – one for each of the newly-created data flows – suggesting that there is no link specified explicitly in the information model which conveys the specific data flow. The results are also communicated to the user with a blue "i" indication on top of each data flow in the diagram. We can follow this design tip to include more comprehensive design details in our model, by specifying each of the existing links as the conveyer of the data flow. Figure 7 shows this option: the user can specify a *Conveying Link* using the *Properties* panel (below the diagram) for the selected data flow (selected in the diagram), which would open a drop-down menu with the links available within our assessment's information model. The conveying link in the specific case (of the data flows defined) can be: 1) a link utilizing the *4G/5G* link from *Shore control centre* to *Communication network* and the *4G/5G* link from *Communication network* to *Ship systems* (specifically *Connectivity Manager*); *or* 2) a link utilizing the *IPv6* connection from *Shore control centre* to *Communication network* and the *4G/5G* link from *Communication network* to *Ship systems*. An elaboration of the connectivity may reveal that several data items are routed using the first option while others are routed using the second one. The original report does not explicitly identify the conveying links, which hinders the exact reproduction of the assessment.

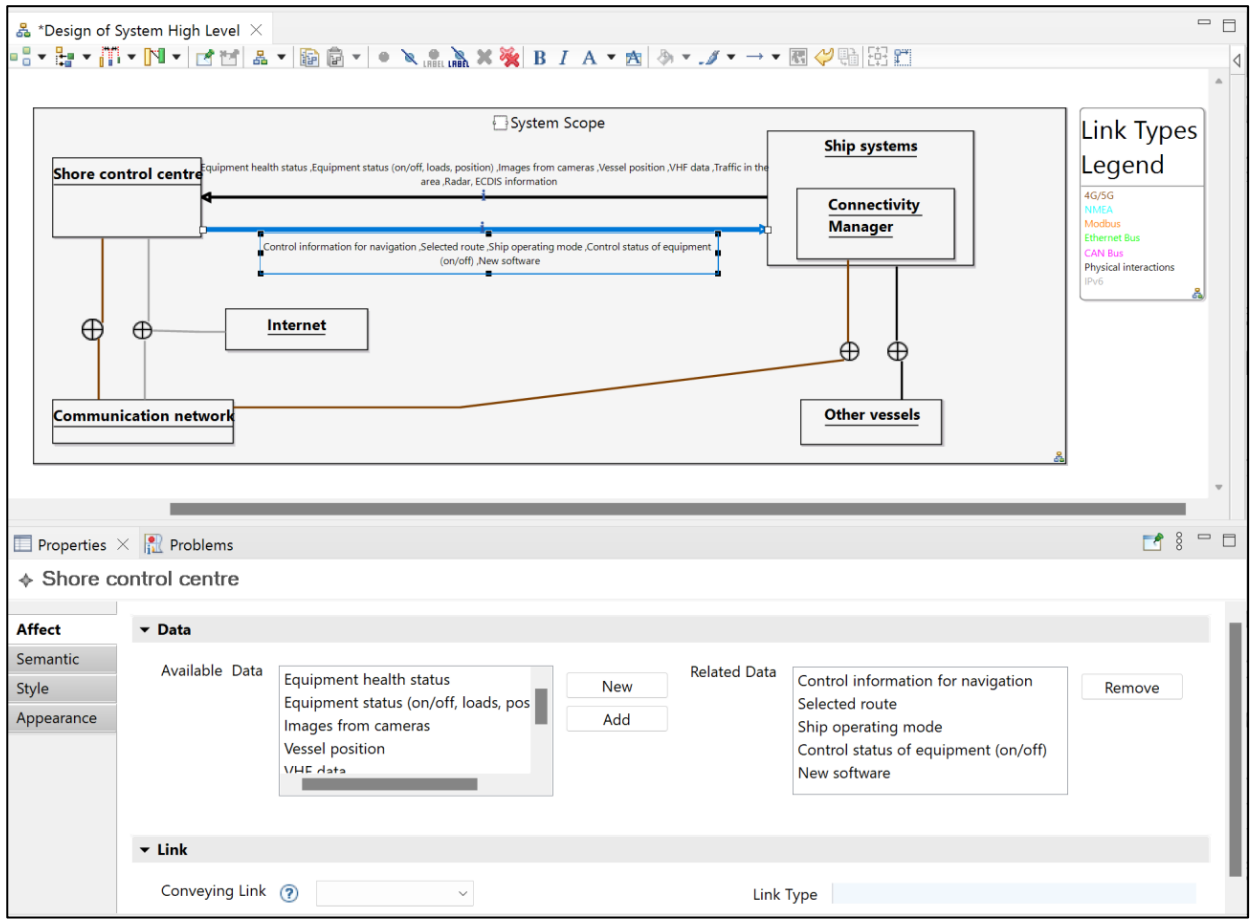

**Figure 7**: **A data flow and its properties, within the TRADES Tool modelling environment.**

The assessment continues by employing cybersecurity domain knowledge, which extends beyond the scope of this article. Briefly, in the remaining Phase A steps, the assessment identifies *terrorists* as the main attack group and vulnerabilities for different components are identified. This leads to the identification of 52 attack scenarios in Phase B (of which only a subset is revealed, later in Phase C as shortly discussed). We note that TRADES misses the ability to explicitly accommodate the attacker's identity; yet this does not have any effect on the rest of the reproduction.

In Phase C of the assessment, the scenarios are ranked with respect to their likelihood and consequence. While the domain-specific ranking mechanisms are beyond the scope of this article, we focus on the resulting likelihood and consequences metrics, FI (frequency index) and SI (severity index) respectively. The assessment employs 7 FI levels and 4 SI levels. It also assigns an integrated Risk Index (RI) to each combination of FI and SI. Then, three levels of risk intolerance are defined: *Low* for RI smaller than 5, *Medium* for RI between 5 and 7

(including), and *High* for RI higher than 7. The original report's Table 9 – shown in Figure 8 – communicates the risk definitions in the popular form of a risk matrix.

**Table 9**
The risk matrix (IMO, 2018).

| FI | Frequency | Severity (SI) | | | |
|---|---|---|---|---|---|
| | | 1 Minor | 2 Significant | 3 Severe | 4 Catastrophic |
| 7 | Frequent | (H) 8 | (H) 9 | (H) 10 | (H) 11 |
| 6 | | (M) 7 | (H) 8 | (H) 9 | (H) 10 |
| 5 | Reasonably probable | (M) 6 | (M) 7 | (H) 8 | (H) 9 |
| 4 | | (M) 5 | (M) 6 | (M) 7 | (H) 8 |
| 3 | Remote | (L) 4 | (M) 5 | (M) 6 | (M) 7 |
| 2 | | (L) 3 | (L) 4 | (M) 5 | (M) 6 |
| 1 | Extremely remote | (L) 2 | (L) 3 | (L) 4 | (M) 5 |

High (H) = Intolerable Risk  
Medium (M) = Tolerable Risk  
Low (L) = Negligible Risk

**Figure 8**: **The risk matrix definitions and structure as they appears in** (Bolbot et al., 2020) **(originally, Table 9)**

Figure 9 shows the reproduction of the assessment's risk matrix definitions using TRADES's risk management summary representation. This automatically generated representation relies on a scoring system definition within the information model. The scoring system includes two types of ordinal elements: one type representing likelihood related definitions, akin to Bolbot et al.'s ordinal FI frequency levels; and another type representing impact related definitions, akin to Bolbot et al.'s ordinal SI severity levels. Accordingly, our scoring system includes 7 likelihood elements and 4 impact elements. In TRADES, each pairing of a likelihood element and an impact element can be associated with a color setting, denoting the risk intolerance. In accordance with Bolbot et al.'s three levels of risk intolerance, we define three colors to be used in the impact-likelihood pairing: Green for *Low*, Yellow for *Medium* and Red for *High*. The resulting matrix is equivalent to Bolbot et al.'s matrix, for any practical use.

AutoShip Risk Management Summary

| | Extremely Remote [1] | [2] | 3-Remote [3] | [4] | Reasonably probable [5] | Difficulty [6] | Frequent [7] |
|---|---|---|---|---|---|---|---|
| Minor [1] | 🟩 | 🟩 | 🟩 | 🟨 | 🟨 | 🟥 | 🟥 |
| Significant [2] | 🟩 | 🟨 | 🟨 | 🟨 | 🟥 | 🟥 | 🟥 |
| Severe [3] | 🟩 | 🟨 | 🟨 | 🟥 | 🟥 | 🟥 | 🟥 |
| Catastrophic [4] | 🟨 | 🟨 | 🟨 | 🟥 | 🟥 | 🟥 | 🟥 |

**Figure 9**: **The risk matrix definitions and structure reproduced using TRADES Tool**

The original report continues by ranking the scenarios, and disclosing those that are rated as *High* risk. Each of these four scenarios is attributed to a specific system. Table 1 offers a simplified view of the relevant table offered in the original report – communicating each of the scenarios (identified by the original unique ID), the respective system under threat, and the risk-related indices FI, SI, RI – as they appear in the original report. We note that Bolbot et al. do not use the risk matrix (Figure 8) for representing the concrete risk scenarios.

**Table 1: Identification of risk scenarios and their scores, as they appear in** (Bolbot et al., 2020) **(originally, Table 12).**

| # (ID) | Scenario | System | FI | SI | RI |
|---|---|---|---|---|---|
| 1 | Combination of social engineering with malware installation | Shore control centre | 5 | 4 | 9 |
| 2 | Getting access to the shore control centre | Shore control centre | 5 | 4 | 9 |
| 10 | Physical attack | Ship control station | 5 | 4 | 9 |
| 7 | Malware installation | Connectivity manager | 4 | 3 | 8 |

We reproduce the scenarios ranking by using the TRADES threat analysis representation, as Figure 10 illustrates. The representation communicates the hierarchical composition of components and the data flows, in synchronization with the information model, and as specified using the design representation (Figure 6). We filtered the representation to show only high-level systems as well as the lower-level components that are assigned with a high-risk scenario. We detail each of the four scenarios as a threat element, and assign each of these to the designated component, based on the information provided in the original report (Table 1). In TRADES, the threat elements appear as external to the system (outside system boundaries), to differentiate them from the system design aspects. We also use the FI and SI ratings from the original report to specify the feasibility and impact of each scenario allocation, using the *Properties* panel. The figure specifically shows the *Catastrophic [4]* SI value assigned as the *Impact* of the concrete risk scenario #1; as well as the *Reasonably probable [5]* FI value (from the previously defined FI levels) being assigned to the same scenario.

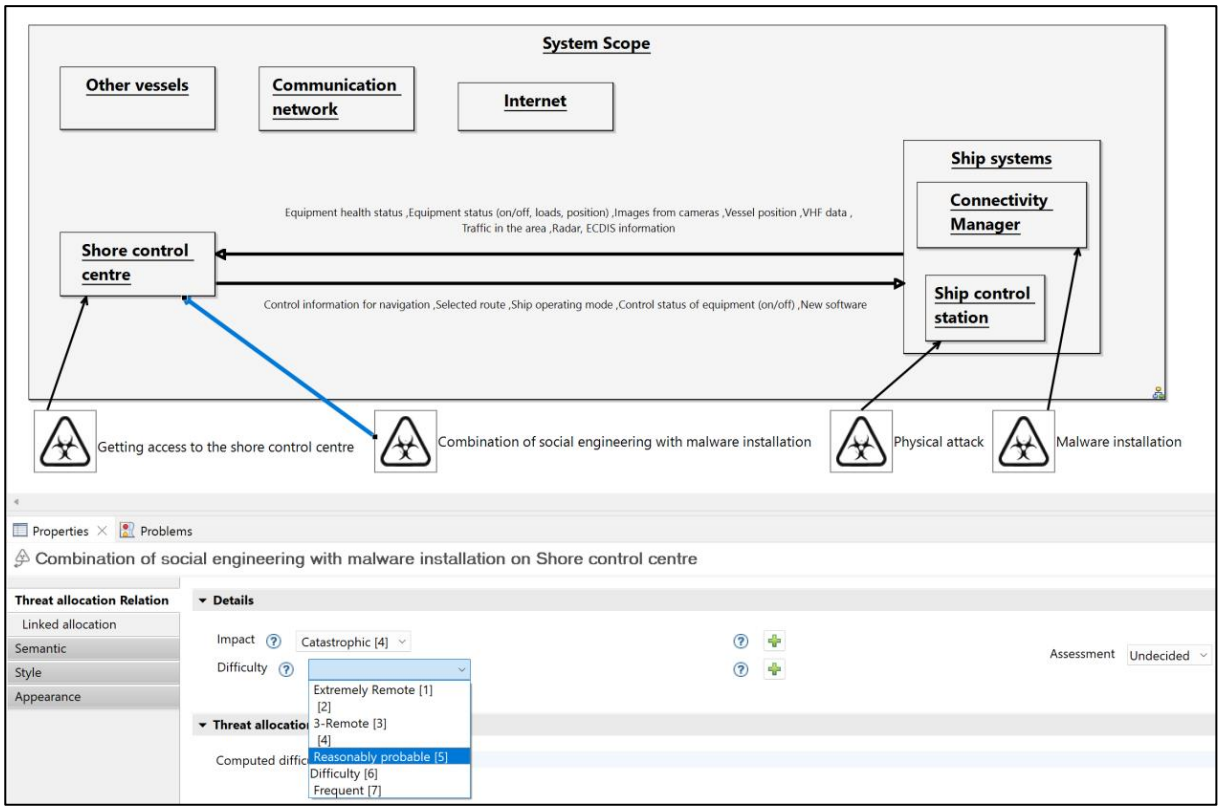

**Figure 10**: **System threat analysis using TRADES Tool**

Figure 11 shows the resulting risk matrix, which is automatically and dynamically generated according to the scenarios' ratings. The figure communicates that one of the scenarios is not within the "red" zone of the matrix, which denotes the high-risk scenarios. We are able to trace that to a wrong value in the scenarios ratings table (Table 1): according to the meta-level risk definitions captured in Figure 8, a scenario with FI=4 and SI=3 should result in RI=7, and not RI=8 as specified in Table 1. While we believe this is an innocent typo, our lived experience shows that such typos are common, especially as the risk assessment incorporates more information and becomes more complicated; and this may result in efforts to address risks that are already sufficiently addressed. The ability to identify such mistakes easily by our model-based representation attests to the usefulness of such approaches to communicate information and as reliable, decision support mechanisms.

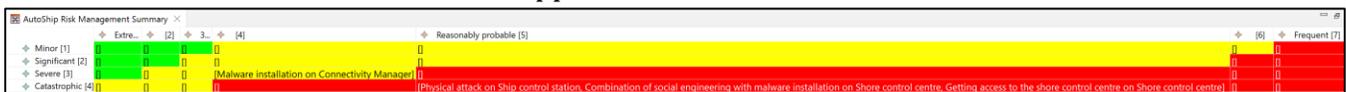

**Figure 11**: **TRADES Tool risk matrix populated with the risk scenarios, as analyzed**

The original report continues with Phase D, suggesting enhancements to the system. Since our reproduction effort focuses on the information aspects and avoids discussion with respect to technical cyber security mechanisms, we continue by reflecting on the updated risk assessment that results from these enhancements.

The original report does not explicitly explain how the security controls mitigate specific scenarios. This presents a gap in assessing whether the controls are relevant and whether the high-risk scenarios are all addressed by the suggested controls. As an illustration, the original report proposes enhancements to several critical components – "*autonomous ship controller, intrusion detection system and navigation system*" – suggesting "*that they operate in a kernel function, so that no software is installed without permission.*" We can safely deduce – based on domain knowledge – that this suggested enhancement may be considered as a mitigation for the *Malware installation* threat. However, this specific enhancement mechanism is not suggested for the *Connectivity Manager* component, which is the actual component subject to the threat in the pertinent high-risk scenario – scenario #7 (Table 1).

Figure 12 illustrates how the TRADES modelling environment can accommodate the enhancement proposition and support the corresponding analysis. The control element *Operate in a kernel function* is now specified as a functional allocation to the *Autonomous Ship Controller* component (now unhidden in the diagram). In TRADES Tool, this allocation is denoted using a shield-like symbol (signifying the control function element) within the boundaries of the specific component. A green mitigation relation arrow from the control element to the threat element (*Malware installation*) denotes the potential applicability of this control to the threat. We also added a concrete risk scenario of applying *Malware installation* to the *Autonomous Ship Controller*, denoted with an arrow from *Malware installation* to *Autonomous Ship Controller*. The listed control mitigates the newly-added scenario, and, therefore, it has a mitigation relation to the specific scenario. Accordingly, the specific scenario (represented by the arrow from the *Malware installation threat* to the *Autonomous Ship Controller* component) is now at an acceptable level, and therefore colored green. According to the original report, the *Malware installation on Connectivity Manager* risk scenario should also be at an acceptable level, and therefore it also appears as a green arrow.

When performing the built-in TRADES diagram validation, a message appears in the *Problems* panel and a blue "i" symbol is indicated on the specific *Malware installation on Connectivity Manager* scenario edge, warning the user that the scenario (TRADES' *threat allocation*) is marked as *accepted* yet does not have any mitigations designed (i.e., included in the model) to support that. This is in contrast with the other accepted scenario (*Malware installation on Autonomous Ship Controller*), which does have a specific control associated with it. The figure therefore explicitly communicates two aspects of the suggested control/enhancement: 1) a control of a certain type is in principle effective against a type of threats; and 2) the specific control (allocated to a specific system level/component) is deemed effective in mitigating a specific risk scenario. The original report only relates to the former aspect, to some extent, yet does not provide for the latter aspect. As a direct result of this gap in the original report, the effect of the suggested enhancements on specific risk scenarios is hard to assess.

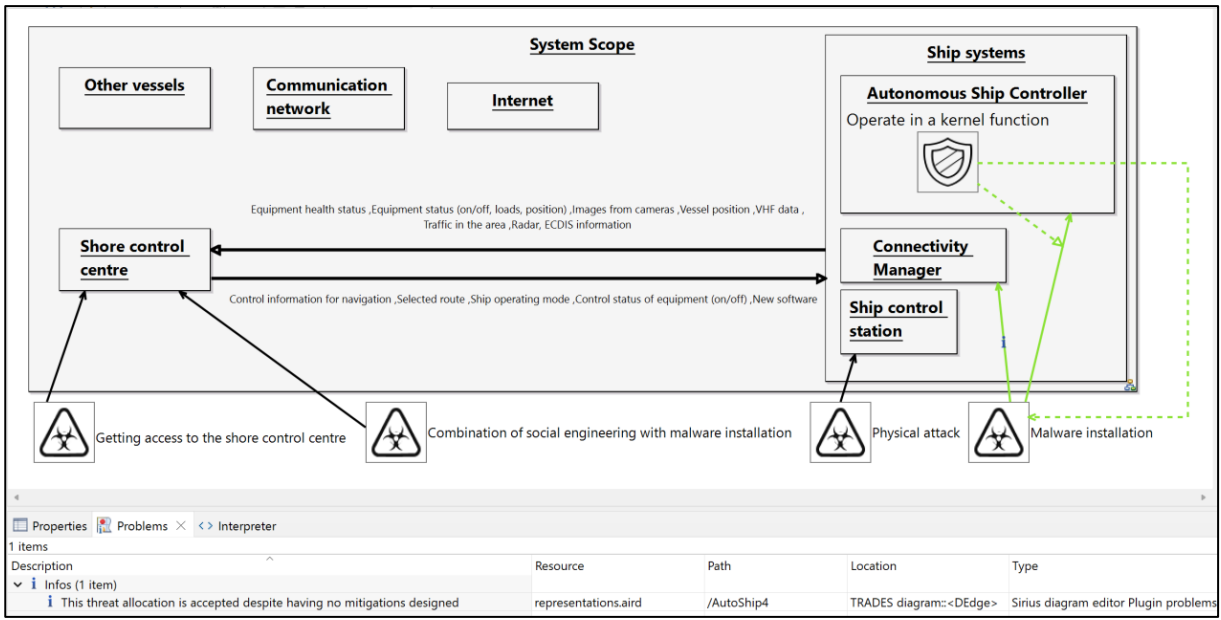

**Figure 12**: **TRADES system threat analysis, with the addition of a control element**

Ignoring the gap in explicitly stating the effect of enhancements on specific risk scenarios, we continue by relying on the original report's change in risk-related indices due to the enhanced design. Bolbot et al. offer updated FI, SI and RI values for the high-risk scenarios. These appear in Table 2. The suggested enhancements – such as Firewalls, Intrusion Detection, Encryption and Authentication – aim to decrease the likelihood of attack. As such, they are expected to affect only the FI index. By comparing Table 2 with Table 1, this is indeed the case, matching our expectation; with only one exception: the SI for Scenario 7 has now increased from 3 to 4. Such an increase is unlikely to occur due to incorporating the suggested enhancements, and this strengthens our position that the original report's Table 12 – specifically the pertinent subset represented here in Table 1 – has an innocent typo. Figure 13 shows the results of updating the SI and FI values, using the TRADES risk management summary representation. All risk scenarios are explicitly situated outside of the high-risk area (denoted in the risk matrix using the red background), as we expect.

**Table 2: Identification of risk scenarios and their scores after suggested mitigation, as they appear in** (Bolbot et al., 2020) **(originally, Table 12).**

| # (ID) | Scenario | System | FI | SI | RI |
|---|---|---|---|---|---|
| 1 | Combination of social engineering with malware installation | Shore control centre | 2 | 4 | 6 |
| 2 | Getting access to the shore control centre | Shore control centre | 2 | 4 | 6 |
| 10 | Physical attack | Ship control station | 1 | 4 | 5 |
| 7 | Malware installation | Connectivity manager | 1 | 4 | 5 |

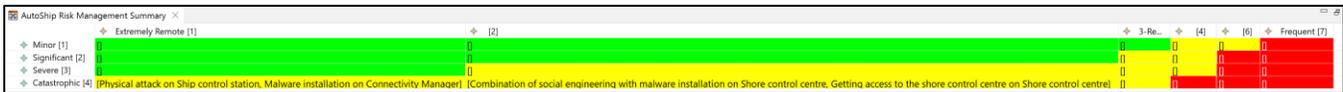

**Figure 13**: TRADES Tool risk matrix populated with the risk scenarios, after mitigation

## 3 Discussion

Security risk assessment is crucial in establishing the security posture of systems, improving their trustworthiness and supporting decision making with respect to the development and the operation of systems. Accordingly, we expect security risk assessments – by both industrial and academic practitioners – to be performed in a systematic, reproducible manner, attesting to high process maturity and enabling consistent decision making. In this article, we use an existing domain-specific, model-based tool to reproduce a prominent security risk assessment publication (Bolbot et al., 2020), focusing on method and information related aspects. In addition to examining aspects of the original report's reproducibility, this effort provides a qualitative benchmark between the free form "pen and paper" approach of the original report with a computer aided design approach, as both are applied to the same real-world scenario, using the same assessment method.

In our reproduction, we deliberately avoid discussions relating to the analyzed cybersecurity posture and to the effectiveness of the suggested enhancements. Instead, we focus on the availability and consistency of information as well as on meta-level mechanisms that are relevant to engineering and risk management. These mechanisms are typically more stable and less temporal, compared with the cybersecurity domain's specifics. As an example, we avoid discussion regarding the effectiveness of requiring software to "*operate in a kernel function*" (one of the enhancements suggested in the original report) for mitigating malware installation; but we do point out that a specific malware installation risk is left unaddressed by this suggested enhancement. Similarly, we avoid discussing whether an attacker in the form of a terrorist has "*technological level 4*" capabilities to perform attacks – as Bolbot et al. assess; but we point out a discrepancy when a specific risk is being considered high-risk whereas according to its ordinal level scores it should be considered medium-risk.

In our reproduction effort, we identified the following issues with respect to the details found in the original report:

1. There is a potential inconsistency regarding the representation of links between the ship and other vessels in the original report (Figure 1, resolved with a specific interpretation in Figure 2).
2. There is a potential inconsistency in the scope of analysis as it unfolded. Specifically, a connection to the Internet, which was not initially disclosed, was introduced during Phase A (Figure 1 vs. Figure 3, and Figure 2 vs. Figure 5).
3. The original report does not allow for immediate reproduction of data flows, due to missing source/destination of received/sent data items (Figure 4, resolved in Figure 6).
4. Data flows are not explicitly associated with the links that convey them (Figure 6). The explicit specification of the conveying links could help to establish a clearer connection between the original schematic diagram (Figure 1, reproduced in Figure 5) with the risk scenarios (represented in Figure 10).
5. The risk ranking of one risk scenario is inconsistent with the report's focus on high-risk scenarios (RI>7).
6. The relevance of proposed enhancements to the high-risk scenarios is not well established.
7. The change in risk-related indices of the risk scenarios could not be corroborated by the disclosed information.

The specific model-based approach that we employed – TRADES – was instrumental in introducing rigor to the reproduction and in the analysis itself. Specifically:

1. The TRADES information model is an instantiation of the TRADES metamodel, and, as such, the construction of the model is constrained to comply with the metamodel (a "correct-by-construction" approach). Therefore, any missing information – such as the missing end of data flows – could be easily identified while constructing the model.
2. TRADES representations are automatically and dynamically synchronized with the information model as their authoritative source of truth, allowing to maintain consistency between views and to identify potential inconsistencies within the original report, such as those relating to the risk indices of one specific risk scenario. Our ability to identify issues easily using model-based representations attests to the usefulness of representations when coupled with information models to communicate information. Here, we particularly established the functionality of the TRADES risk summary representation as effective. This representation is basically a model-based version of the risk matrix – a prominent representation of risks extensively used in risk management. Communicating risk scenarios within the context of the risk matrix (using color-coded risk intolerance as a heat map), revealed an inconsistency. Had this type of representation been used in the original report for populating the risk scenarios (and not just for defining the risk scoring system), the specific inconsistency could have been easily identified and prevented. Furthermore, the TRADES risk summary representation is effective in representing the security posture of the original design vs. the suggested enhanced design (Figure 11 vs Figure 13), offering a communicable artifact that the original report misses.
3. The TRADES Tool's built-in validation rules – querying the information model and representations with respect to the metamodel – facilitated our identification of missing information and of potential gaps or enhancements in the systematic application of the original report's cybersecurity risk assessment method. An example of this is the missing mitigation attribution between suggested enhancements to the scenarios they mitigate. This is a conceptual gap which we also encounter in other risk assessments and security designs. To our knowledge, this is the first report utilizing the TRADES Tool validation rules capability.

Using TRADES successfully and naturally to reproduce an external, independently performed risk assessment provides further validation of the methodology and tool, extending previously reported cases (Shaked, 2023) and showcasing new features for detailing the architecture of the system and validation. Specifically, using TRADES without any significant adaptations attests to the correct identification of its underlying domain ontology. The few exceptions that we identified – and might be addressed via future research – are:
1. TRADES misses the ability to accommodate the attacker's identity. This did not have any direct consequence on the ability to reproduce the original report, as the attacker's identity is simply translated into the risk scenarios and the tolerance levels associated with the risks (specifically, the frequency index). However, we identify that explicitly mentioning the attacker and its capabilities may provide additional risk assessment functionality, especially if well integrated with other model elements, such as the impact and feasibility levels. Both Yarcaf (Ekstedt et al., 2023) and ROSE (Oliveira et al., 2022) include the concept of *Attacker* in the security-oriented ontology, so adding this concept to TRADES might provide for a better alignment with some of the domain's prominent ontologies.
2. TRADES does not support directionality for links. According to previous TRADES publications, this is rooted in cybersecurity domain expertise, in which any link – even if it conveys directional data (which is supported by TRADES) – can present a risk to both ends of the link. We assess that integrating support for directional links in TRADES is easy, e.g., an enumerated attribute can be added to the link concept to indicate its directionality and the design notation can then use this attribute to represent unidirectional links with arrowheads. However, from a conceptual standpoint, this issue requires careful consideration and justification.

Selecting a specific, computer aided methodology for our analysis (TRADES) does not limit the generality of both the approach and the case for using computer aided design methodologies as mechanisms to underpin decision making. Any fit for purpose computer aided design methodology could be useful in improving the rigor of analyzing systems and, consequently, making pertinent, well-informed decisions in a systematic manner. Further research can continue to establish the value of using computer aided methodologies for analyzing the security of various types of systems as well as in additional domains. Future research could also aim to characterize and uncover criteria for the "fit for purpose" requirement of such methodologies. Our research suggests that a key aspect of an effective, "fit for purpose" computer aided methodology is having a codified ontology that is theory-informed as well as empirically grounded in the domain of application and decision making.

## 4 Conclusion

Developing and sustaining systems for security and trustworthiness is both a goal and a challenge in contemporary engineering. Modern system designs are becoming increasingly complex, and their analysis and pertinent decision making pose a significant cognitive load. Computer aided methodologies should, therefore, be considered an enabler of systematic analysis of such designs. In this article, we reproduced – and reflected on – a previous, peer-reviewed security analysis, using a computer aided methodology. Consequently, we established the effectiveness of using computer aided methodologies for analyzing the security of systems' designs and rigorously supporting decision making.

## ACKNOWLEDGMENTS

The author wishes to thank Victor Bolbot, Gerasimos Theotokatos, Evangelos Boulougouris and Dracos Vassalos for their inspiring research article and for making it available as an open access article (CC BY 4.0 license), which made this contribution possible.
The author would also like to express his gratitude to Professor Tom Melham for his generous support.
This work is supported by Innovate UK, grant #75243.